\documentclass[12pt, draftcls, onecolumn]{IEEEtran}

\usepackage{caption}

\ifCLASSINFOpdf
  \usepackage[pdftex]{graphicx}
  \DeclareGraphicsExtensions{.pdf,.jpeg,.png}
\else
  \usepackage[dvips]{graphicx}
\fi
\usepackage{subfigure}
\usepackage{epstopdf}
\usepackage[cmex10]{amsmath}
\usepackage{amssymb}
\usepackage{wasysym}
\usepackage{algorithm}
\usepackage{algorithmic}
\usepackage{array}
\usepackage{cite}
\usepackage{color}
\usepackage{url}
\usepackage{verbatim}
\usepackage{booktabs}
\usepackage{epsfig,latexsym}
\usepackage{subeqnarray}
\usepackage{cases}
\usepackage{cite}
\usepackage{hyperref}
\usepackage{makecell}
\usepackage{amsmath}
\usepackage{stfloats}
\usepackage[justification=centering]{caption} 

\begin{document}
%
\title{Multibeam Satellite Communications with Energy Efficiency Optimization}

\author{Chenhao~Qi, Yang Yang, Rui Ding, Shichao Jin and Dunge Liu
\thanks{Chenhao~Qi and Yang Yang are with the School of Information Science and Engineering, Southeast University, Nanjing 210096, China (Email: qch@seu.edu.cn).}
}

\markboth{}
{Shell \MakeLowercase{\textit{et al.}}: Bare Demo of IEEEtran.cls for Journals}
\maketitle

\begin{abstract}
Energy efficiency (EE) is an important aspect of satellite communications. Different with the existing algorithms that typically use the first-order Taylor lower bound approximation to convert non-convex EE maximization (EEM) problems into convex ones, in this letter a two-step quadratic transformation method is presented. In the first step, the fractional form of the achievable rate over the total power consumption is converted into a non-fractional form based on quadratic transformation. In the second step, the fractional form of the signal power over the interference-and-noise power is further converted into a non-fractional form, still based on quadratic transformation. After the two-step quadratic transformation, the original EEM problem is converted into an equivalent convex one. Then an alternating optimization algorithm is presented to solve it by iteratively performing two stages until a stop condition is satisfied. Simulation results show that the presented algorithm can fast converge and its performance is better than that of the sequential convex approximation algorithm and the multibeam interference mitigation algorithm.

\end{abstract}

\section{Introduction}
Multibeam satellite systems (MSS) and frequency reuse which can meet the growing demand for high-data-rate service, have become one of most suitable choice for the next generation satellite communications~\cite{SDAward2020Beamforming}. Considering that future
wireless networks will be required to support reliable and high-quality user service in remote areas, we are interested in the MSS since it can provide great flexibility in deploying wireless networks without geographical constraints in these areas~\cite{ZLin2020Secure}. To increase the spectral efficiency of the MSS, multiple-input multiple-output antenna arrays, high frequency reuse and efficient satellite precoding can be applied~\cite{Vazquez2015Precoding}. On the other hand, power consumption which has a large impact on the lifetime and quality of the satellite is an innegligible limitation of the MSS. Therefore, we consider another performance metric named as energy efficiency (EE) of the MSS, where the EE is defined as a ratio of the achievable rate over the total power consumption of satellite communications. Note that improving the EE can effectively extend the service time of the satellite as well as reducing its size~\cite{GGiambene2018Satellite}.

Some works have already investigated the EE maximization (EEM) problem of the MSS. In~\cite{qi2018precoding}, under the constraints of total power and quality of service (QoS) for the MSS, two precoding design algorithms based on zero-forcing and sequential convex approximation (SCA) are proposed, where the SCA algorithm first converts the hyperbolic constraints into convex ones with second-order cone representations, and then uses the first-order Taylor lower bound approximation (FTLBA) to convert the joint convex constraints into linear constraints. In~\cite{CNEfrem2020Dynamic}, the EEM problem is first converted to be an optimization problem minimizing the weighted sum of unmet system capacity (USC) and total power consumption, where USC is defined as the difference between the users' requested capacity and the provided capacity; and then the problem is rewritten as the equivalent differentiable epigraph-form so that it can be finally converted to be a convex optimization problem via the FTLBA. In~\cite{CQi2020Energy}, a precoding algorithm for multicast scenario under the constraints of total power and QoS of satellite is proposed, where the EEM problem is first formulated as a concave-convex fractional programming (CCFP) one using the FTLBA and is then reformulated as a convex optimization problem based on Charnes-Cooper transformation.

Different with the above algorithms that all use the FTLBA method to convert non-convex EEM problems of the MSS into convex ones, in this letter we present a two-step quadratic transformation method. In the first step, we convert the fractional form of the achievable rate over the total power consumption into a non-fractional form based on quadratic transformation. In the second step, we further convert the fractional form of the signal power over the interference-and-noise power into a non-fractional form, still based on quadratic transformation. After the two-step quadratic transformation, the original EEM problem is converted to an equivalent convex one. Then we present an alternating optimization algorithm to solve the equivalent convex problem by iteratively performing two stages until a stop condition is satisfied.




Notations: Symbols $[\boldsymbol{M}]_{m,n}$, $[\boldsymbol{v}]_{n}$, $\boldsymbol{I}_{L}$, ${\mathbb{C}}^{M}$, ${\mathbb{R}}^{M}$, $\Re\{\boldsymbol{M}\}$ and $\mathcal{CN}$ represent the entry on the $m$th-row and $n$th-column of a matrix $\boldsymbol{M}$, the $n$th entry of a vector $\boldsymbol{v}$, the $L\times L$ identity matrix, the set of complex-valued column vectors with length $M$, the set of real-valued column vectors with length $M$, the real part of a matrix $\boldsymbol{M}$ and the complex Gaussian distribution, respectively.

\section{System Model}\label{sec.SystemModel}
A satellite communication system considered in this letter is set up with a broadband multibeam satellite serving $K$ users. By the array feed reflector on the satellite, $M$ feeder signals are converted into $K$ transmit signals. There are $K$ beams in total, since the coverage of each one of the $K$ signals on the ground forms a beam. For the unicast scenario, each beam of the satellite can only serve one user within a time slot, where multiple users can be served using time-division multiple access. To improve the spectral efficiency of the precious satellite frequency band, we adopt full frequency reuse.

The downlink channel matrix $\boldsymbol{H} \in \mathbb{C}^{K \times M}$ from the multibeam satellite to the $K$ ground users, can be expressed as~\cite{Christopoulos2015Multicast,ChannelModeling2021}
\begin{equation}\label{ChannelModel}
    \boldsymbol{H} = \boldsymbol{\Phi}\boldsymbol{D}
\end{equation}
where $\boldsymbol{\Phi} \in \mathbb{C}^{K \times K}$ represents the signal phase matrix generated by the different propagation paths from the satellite to the users. Since the distance between the satellite and the user is much longer than that between the adjacent satellite antenna feeds, it is generally assumed that the phases between the user and all antenna feeds are the same in the line-of-sight (LOS) propagation environment~\cite{qi2018precoding}. Therefore, $\boldsymbol{\Phi}$ is a diagonal matrix with diagonal entries defined as $[\boldsymbol{\Phi}]_{i,i} = e^{j\phi_i},i=1,2,\ldots,K$, where $\phi_i$ obeys the uniform distribution in $(0,2\pi)$. Except the diagonal entries, the other entries of $\boldsymbol{\Phi}$ are all zero, i.e., $[\boldsymbol{\Phi}]_{i,l} = 0$ for $i \neq l$. Note that $\boldsymbol{D} \in \mathbb{R}^{K \times M}$ in~\eqref{ChannelModel} represents the multibeam antenna pattern, which contains the radiation pattern of the satellite, the gain of the receiving antenna, the loss of the propagation path and the power of various noise. The entry on the $k$th-row and $m$th-column of $\boldsymbol{D}$, for $k=1,2,\ldots,K$ and $m=1,2,\ldots,M$, is given by
\begin{equation}\label{AntennaPatternD}
    [\boldsymbol{D}]_{k,m} = \frac{\sqrt{G_RG_{k,m}}}{4\pi\frac{d_k}{\lambda}\sqrt{\kappa T_R B_W}}
\end{equation}
where $G_R$ and $G_{k,m}$ are the gain of the receiving antenna at the user terminal and the gain between the $m$th feeder on the satellite and the $k$th user. $d_k$ represents the LOS distance between the $k$th user and the satellite. Besides, $\lambda, \kappa, B_W$ and $T_R$ represent the wavelength, Boltzmann's constant, bandwidth and receiving noise temperature under clear sky environment, respectively.

We denote the data vector sent by the multibeam satellite to all $K$ users as  
\begin{equation}
	\boldsymbol{s} \triangleq [s_1,s_2,\ldots,s_K]^T\in \mathbb{C}^{K}
\end{equation}
where we assume that $\boldsymbol{s}\sim\mathcal{CN}(\boldsymbol{0},\boldsymbol{I}_K)$ without loss of generality. To mitigate the inter-beam interference caused by the satellite channels, a common preprocessing technique is to introduce a precoding matrix to combat the channel distortion. We denote the precoding matrix to be designed as
\begin{equation}
	\boldsymbol{W}\triangleq [\boldsymbol{w}_1,\boldsymbol{w}_2,\ldots,\boldsymbol{w}_K]\in\mathbb{C}^{M \times K}.
\end{equation}
The transmitted signal of the multibeam satellite can be written as
\begin{equation}
    \boldsymbol{x} = \boldsymbol{W}\boldsymbol{s}.
\end{equation}

Then the received signal can be expressed as
\begin{equation}
    \boldsymbol{y} = \boldsymbol{H} \boldsymbol{x} + \boldsymbol{n}
\end{equation}
where $\boldsymbol{y}\triangleq [y_1,y_2,\ldots,y_K]^T\in \mathbb{C}^{K }$ is the signal vector received by all $K$ users, and $\boldsymbol{n} \triangleq [n_1,n_2,\ldots,n_K]^T\in \mathbb{C}^{K}$ is an additive white Gaussian noise (AWGN) vector with each entry independent and identically distributed, i.e., $\boldsymbol{n} \sim \mathcal{CN}(\boldsymbol{0},\sigma^2\boldsymbol{I}_K)$.

We further define 
\begin{equation}
	\boldsymbol{H} \triangleq [\boldsymbol{h}_1^T,\boldsymbol{h}_2^T,\ldots,\boldsymbol{h}_K^T ]^T
\end{equation}
where $\boldsymbol{h}_k\in\mathbb{C}^{1 \times M}$ denotes the $k$th row of $\boldsymbol{H}$. Therefore, we can express the received signal by the $k$th user as
\begin{equation}
    y_k = \boldsymbol{h}_k \boldsymbol{w}_ks_k + \sum_{l \in \mathcal{K}, l \neq k}{\boldsymbol{h}_k \boldsymbol{w}_ls_l} + n_k,~k \in \mathcal{K}
\end{equation}
where $\mathcal{K}\triangleq\{1,2,\ldots,K\}$ represents the user index set. The signal-to-interference-and-noise ratio
(SINR) and achievable rate $R_k(\boldsymbol{W})$ of the $k$th user can be expressed respectively as
\begin{equation}\label{Gamma_k}
    {\Gamma}_k \triangleq \frac{|\boldsymbol{h}_k\boldsymbol{w}_k|^2}{\sum_{l \in \mathcal{K} \atop l\neq k} {\boldsymbol{h}_k\boldsymbol{w}_l\boldsymbol{w}_l^H\boldsymbol{h}_k^H} +\sigma^2},
\end{equation}
and
\begin{equation}\label{R_k_W}
    R_k(\boldsymbol{W}) \triangleq \log{(1+{\Gamma}_k)}.
\end{equation}

Now the EEM problem of the MSS can be formulated as
\begin{subequations}\label{EEM0}
    \begin{align}
        \max_{\boldsymbol{W}}&~\frac{\sum_{k \in \mathcal{K}} \alpha_k R_k(\boldsymbol{W})}{\sum_{k \in \mathcal{K}} \|\boldsymbol{w}_k\|_2^2 + P_0} \label{EEMObj}\\
        \text{s.t.}~&\sum_{k \in \mathcal{K}} \|\boldsymbol{w}_k\|_2^2 \leq P_T \label{TotalPowerConst}\\
        &~\Gamma_k \geq \overline{\Gamma}_k,~ k \in \mathcal{K} \label{QoSconst}
    \end{align}
\end{subequations}
where $\boldsymbol{\alpha} \triangleq \{\alpha_1,\alpha_2,\ldots,\alpha_K\}$ is a set representing the predefined weights of all the $K$ beams, and $ P_T $ represents the maximum transmission power supported by the power amplifiers on the multibeam satellite. Note that the power consumption on the satellite mainly involves the power supply, various circuit blocks, cooling system, and etc, which are independent of $\boldsymbol{W}$ and can be regarded as a constant working power $P_0$.  $\overline{\Gamma}_k$ represents the threshold SINR of the $k$th user and is related to different QoS requirement of different users~\cite{CQi2020Energy}.

\section{Energy Efficient Precoding Design}\label{sec.EEMax}
Note that \eqref{EEM0} is a CCFP problem in a fractional form of the achievable rate over the total power
consumption. In this work, we will present a two-step quadratic transformation method. In the first step, we will convert the aforementioned fractional form into a non-fractional one based on quadratic transformation~\cite{KShen2018Fractional}. First, we introduce a real-valued variable $\mu$ to decouple the numerator and denominator in the objective function of \eqref{EEMObj}. Then \eqref{EEMObj} can be transformed as the following objective
\begin{equation}\label{Decouple}
    \max_{\boldsymbol{W},~\mu} 2\mu \big(\sum_{k \in \mathcal{K}} \alpha_k R_k(\boldsymbol{W})\big)^\frac{1}{2}
    - \mu^2(\sum_{k \in \mathcal{K}} \|\boldsymbol{w}_k\|^2_2 + P_0)
\end{equation}
where the optimal value of $\mu$ for a given $\boldsymbol{W}$ can be easily calculated from a quadratic function as
\begin{equation}\label{VarMu}
    \mu^* = \frac{\sqrt{\sum_{k \in \mathcal{K}} \alpha_k R_k(\boldsymbol{W})}}{\sum_{k \in \mathcal{K}} \|\boldsymbol{w}_k\|^2_2 + P_0}.
\end{equation}


We will show that \eqref{Decouple} is equivalent to \eqref{EEMObj}. For simplicity, the objective function of \eqref{Decouple} is rewritten as
\begin{equation}\label{MaxV}
    v(\boldsymbol{W},\mu) \triangleq 2\mu \sqrt{N(\boldsymbol{W})} - \mu^2 D(\boldsymbol{W})
\end{equation}
where
\begin{equation}
  N(\boldsymbol{W}) \triangleq \sum_{k \in \mathcal{K}} \alpha_k R_k(\boldsymbol{W}),
\end{equation}
and
\begin{equation}
  D(\boldsymbol{W}) \triangleq \sum_{k \in \mathcal{K}} \|\boldsymbol{w}_k\|^2_2 + P_0.
\end{equation}
For such a quadratic function with respect to $\mu$, we can obtain the optimal $\mu^*$ in \eqref{VarMu} by setting 
\begin{equation}
	\frac{\partial v(\boldsymbol{W},\mu) }{ \partial \mu} = 0
\end{equation}
while the maximum of $v(\boldsymbol{W},\mu)$ is 
\begin{equation}
	v(\boldsymbol{W},\mu^*) = \frac{ N(\boldsymbol{W}) }{D(\boldsymbol{W}) } 
\end{equation}
as in \eqref{EEMObj}.

%


It is seen that the challenge of solving \eqref{Decouple} mainly comes from $R_k(\boldsymbol{W})$, since the second term of \eqref{Decouple} is concave and the function of $(\cdot)^{\frac{1}{2}}$ is a monotonically increasing concave function.

Therefore, in the second step, we will convert the fractional form of the signal power over the interference-and-noise power in \eqref{Gamma_k} into a non-fractional one, still based on quadratic transformation. We introduce $K$ complex-valued variables $z_k,~k=1,2,\ldots,K$, just similar as the first step. Then \eqref{Gamma_k} can be rewritten as
\begin{equation}\label{Second_step_Gamma_k}
    {\Gamma}_k = 2\Re\{z_k^H \boldsymbol{h}_k \boldsymbol{w}_k\} -
    z_k^H(\sigma^2 + \sum_{l\in\mathcal{K} \atop l\neq k}{\boldsymbol{h}_k\boldsymbol{w}_l\boldsymbol{w}_l^H\boldsymbol{h}_k^H})z_k
\end{equation}
which is based on the similar derivation from \eqref{EEMObj} to \eqref{Decouple}. Similar to~\eqref{VarMu}, the optimal value of $z_k$ for a given $\boldsymbol{W}$ is
\begin{equation}\label{VarZk}
    z_k^* = \frac{|\boldsymbol{h}_k \boldsymbol{w}_k|}{\sum_{l\in\mathcal{K} \atop l\neq k}{\boldsymbol{h}_k\boldsymbol{w}_l\boldsymbol{w}_l^H\boldsymbol{h}_k^H} + \sigma^2}.
\end{equation}
We define $\boldsymbol{z} \triangleq [z_1,z_2,\ldots,z_K]$. Then \eqref{Decouple} can be rewritten as
\begin{subequations} \label{EEM_Mu}
    \begin{align}
        \max_{\boldsymbol{W},\mu,\boldsymbol{z}} &~ 2\mu ~\Big(\sum_{k=1}^K \alpha_k \log\big(1 +  2\Re\{z_k^H \boldsymbol{h}_k \boldsymbol{w}_k\} -z_k^H(\sigma^2 + \sum_{l\in\mathcal{K} \atop l\neq k}{\boldsymbol{h}_k\boldsymbol{w}_l\boldsymbol{w}_l^H\boldsymbol{h}_k^H})z_k   \big) \Big)^{\frac{1}{2}} -\mu^2(\sum_{k=1}^K{\|\boldsymbol{w}_k\|_2^2} + P_0)
        \\
        \text{s.t.}~&\sum_{k=1}^K \|\boldsymbol{w}_k\|_2^2 \leq P_T \\
        &~2\Re\{z_k^H \boldsymbol{h}_k \boldsymbol{w}_k\} -
    z_k^H(\sigma^2 + \sum_{l\in\mathcal{K} \atop l\neq k}{\boldsymbol{h}_k\boldsymbol{w}_l\boldsymbol{w}_l^H\boldsymbol{h}_k^H})z_k \geq \overline{\Gamma}_k.
    \end{align}
\end{subequations}


To efficiently solve the convex optimization problem~\eqref{EEM_Mu}, we may resort to the alternating optimization. Since \eqref{EEM_Mu} involves the optimization of $\boldsymbol{W},\mu,$ and $\boldsymbol{z}$, the alternating optimization includes the following two stages:
 \begin{enumerate}
\item In the first stage, given $\boldsymbol{W}$, we determine $\mu$ and $z_k,~k=1,2,\ldots,K$ as follows. We update $\mu$ by
\begin{equation}\label{Update_mu}
  \mu \leftarrow \mu^*
\end{equation}
where $\mu^*$ comes from \eqref{VarMu}. We update $z_k,~k=1,2,\ldots,K$ by
\begin{equation}\label{Update_z}
  z_k \leftarrow z_k^*
\end{equation}
where $z_k^*$ comes from \eqref{VarZk}.

\item In the second stage, given $\mu$ and $z_k,~k=1,2,\ldots,K$, we determine $\boldsymbol{W}$ as follows. We introduce $K$ real-valued variables $\beta_k,k=1,2,\ldots,K$ satisfying
    \begin{equation}\label{constraint_beta}
      \beta_k \geq \sum_{l\in\mathcal{K} \atop l \neq k}{\boldsymbol{h}_k\boldsymbol{w}_l\boldsymbol{w}_l^H\boldsymbol{h}_k^H}+\sigma^2.
    \end{equation}
    Then we introduce $K$ real-valued variables $\gamma_k,k=1,2,\ldots,K$ satisfying
    \begin{equation} \label{constraint_gamma_k_1}
      \gamma_k \geq \overline{\Gamma}_k,
    \end{equation}
and
\begin{equation}\label{constraint_gamma_k_2}
   \gamma_k \leq 2\Re\{z_k^H \boldsymbol{h}_k \boldsymbol{w}_k\} -
    z_k^H(\sigma^2 + \sum_{l\in\mathcal{K} \atop l\neq k}{\boldsymbol{h}_k\boldsymbol{w}_l\boldsymbol{w}_l^H\boldsymbol{h}_k^H})z_k.
\end{equation}
In fact, \eqref{constraint_gamma_k_2} can be rewritten as
\begin{equation}\label{constraint_gamma_k_22}
  \gamma_k \leq 2\Re\{z_k^H\boldsymbol{h}_k\boldsymbol{w}_k\}-z_k^H\beta_kz_k.
\end{equation}
According to \eqref{R_k_W}, \eqref{Second_step_Gamma_k} and \eqref{constraint_gamma_k_2}, we have
\begin{equation}\label{constraint_R_k_W}
  R_k(\boldsymbol{W}) \geq \log{(1+\gamma_k)}.
\end{equation}

With these settings, \eqref{EEM_Mu} can be further converted as
\begin{subequations}\label{EEM_Fin}
    \begin{align}
        \max_{\boldsymbol{W},\boldsymbol{\beta},\boldsymbol{\gamma}}&~2\mu\big(\sum_{k\in\mathcal{K}}{\alpha_k R_k(\boldsymbol{W})}\big)^{1/2}
        -\mu^2(\sum_{k\in\mathcal{K}}{\|\boldsymbol{w}_k\|_2^2+P_0}) \label{f_k}\\
        ~\text{s.t.}~~&\sum_{k\in\mathcal{K}} \|\boldsymbol{w}_k\|_2^2 \leq P_T \\
        &~\eqref{constraint_beta},\eqref{constraint_gamma_k_1}, \eqref{constraint_gamma_k_22}, \eqref{constraint_R_k_W}.
    \end{align}
\end{subequations}
which can be solved by the existing convex optimization toolbox such as the CVX.
\end{enumerate}

\begin{algorithm}[!t]
	\caption{Alternative Optimization for Energy Efficient Precoding Design}
	\label{precoding_algorithm}
	\begin{algorithmic}[1]
		\STATE \emph{Input:} $\boldsymbol{H}$, $\sigma^2$, $\overline{\Gamma}_k$, $P_T$, $P_0$, $\xi$.
		\STATE \emph{Initialization:} $t \leftarrow 0$.
		\STATE Initialize $\boldsymbol{W}$ randomly via \eqref{TotalPowerConst} and \eqref{InitializeW_2}.
		\REPEAT
		\STATE Update $\mu$ and $\boldsymbol{z}$ via \eqref{Update_mu} and \eqref{Update_z}, respectively.
		\STATE Update $\boldsymbol{W}$ by solving \eqref{EEM_Fin}.
		\STATE $t \leftarrow t + 1$.
		\UNTIL \textit{stop condition} is satisfied.
		\STATE \emph{Output:} $\boldsymbol{W}$.
	\end{algorithmic}
\end{algorithm}

We iteratively perform the above two stages until a \textit{stop condition} is satisfied. We might simply set the \textit{stop condition} as the maximum number of iterations being reached. Note that sometimes it might be difficult to determine the maximum number of iterations, since the small value of it leads to unconverged performance and the large value of it leads to wasted computational resource. Therefore, it might be better set as
\begin{equation}\label{StopCondition}
    \Big|\sum_{k\in\mathcal{K}}(f_k^{(t)}-f_k^{(t-1)})\Big| \leq \xi
\end{equation}
where $f_k^{(t)}$ denotes the value of the objective function of $\eqref{f_k}$ at the $t$th iteration and $\xi$ is a predefined convergence threshold. Smaller $\xi$ leads to slower convergence but better performance. We normally set $\xi = 10^{-3}$.

The procedure of the alternating optimization algorithm for the energy efficient precoding design is summarized in \textbf{Algorithm~1}. To initialize $\boldsymbol{W}$ for the first iteration, i.e., $t=0$, of the algorithm, we randomly generate feasible solutions of $\boldsymbol{W}$ satisfying \eqref{TotalPowerConst} and
\begin{equation}\label{InitializeW_2}
  \frac{|\boldsymbol{h}_k\boldsymbol{w}_k|^2}{\sum_{l \in \mathcal{K} \atop l\neq k} {\boldsymbol{h}_k\boldsymbol{w}_l\boldsymbol{w}_l^H\boldsymbol{h}_k^H} +\sigma^2} \geq  \overline{\Gamma}_k,~k=1,2,\ldots,K.
\end{equation}

\section{Numerical Results}\label{sec.NumericalResult}
The single-feed single-beam structure is adopted to simulate a multibeam satellite communication system operating in a synchronous orbit and working in 20GHz Ka band. We consider $K=8$ ground users and use $M=8$ feeds on the satellite to generate 8 beams. For simplicity, we set $\alpha_1=\alpha_2=\cdots=\alpha_K=1$. The parameters setting is listed in Table~\ref{parameters_table}. The feed radiation pattern involved in our simulation is from the European Space Agency (ESA)~\cite{qi2018precoding}. Besides, we set the noise power $\sigma^2=1$ normalized by $\kappa T_R B_W$ in~\eqref{AntennaPatternD}. The Boltzmann constant is denoted as $\kappa=1.38 \times 10^{-23}$~J/K. To evaluate our algorithm, we include the performance comparisons with the SCA algorithm~\cite{qi2018precoding} and multibeam interference mitigation (MBIM) algorithm~\cite{Joroughi2017Generalized}, all in the unicast scenario of satellite communications.
\begin{table}[h]
\centering
\caption{Parameters Setting}\label{parameters_table}
\begin{tabular}{c|c}
\Xhline{1 pt}
Parameter & Value \\ \hline
Carrier frequency & 20 GHz (Ka band) \\
Satellite height &  35786 km\\
User antenna gain & 41.7 dBi \\
$G/T$ & 17.68 dB/K \\
Total bandwidth ($B_W$) & 500 MHz  \\
\Xhline{1 pt}
\end{tabular}
\end{table}

\begin{figure}[!t]
\centering
\includegraphics[width=90mm]{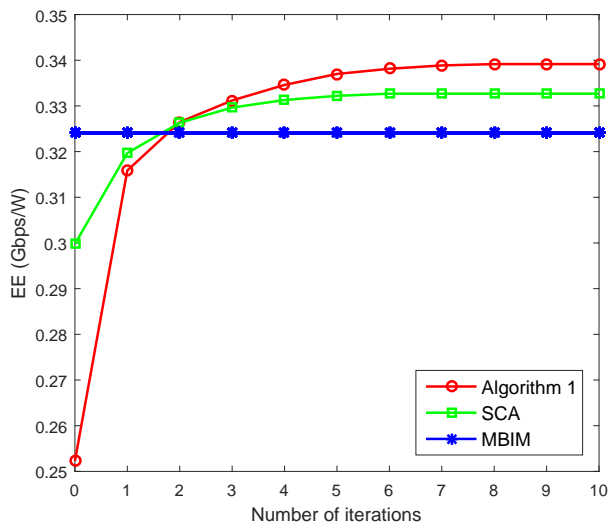}
\caption{Convergence of three algorithms.}
\label{Iteration}
\end{figure}

Fig. \ref{Iteration} compares the convergence of these three algorithms. We set $P_T=10$dBW and $P_0=10$dBW~\cite{CQi2020Energy}. It can be observed that the starting point of \textbf{Algorithm~1} at the first iteration is much worse than the other two algorithms. However, only after one iteration, the performance of \textbf{Algorithm~1} is very close to that of the other two algorithms; and after two iterations, \textbf{Algorithm~1} even outperforms the other two. It is shown that \textbf{Algorithm~1} converges in 7 iterations and can achieve around 2\% and 5\% performance improvement over the SCA and MBIM, respectively.

\begin{figure}[!t]
\centering
\includegraphics[width=90mm]{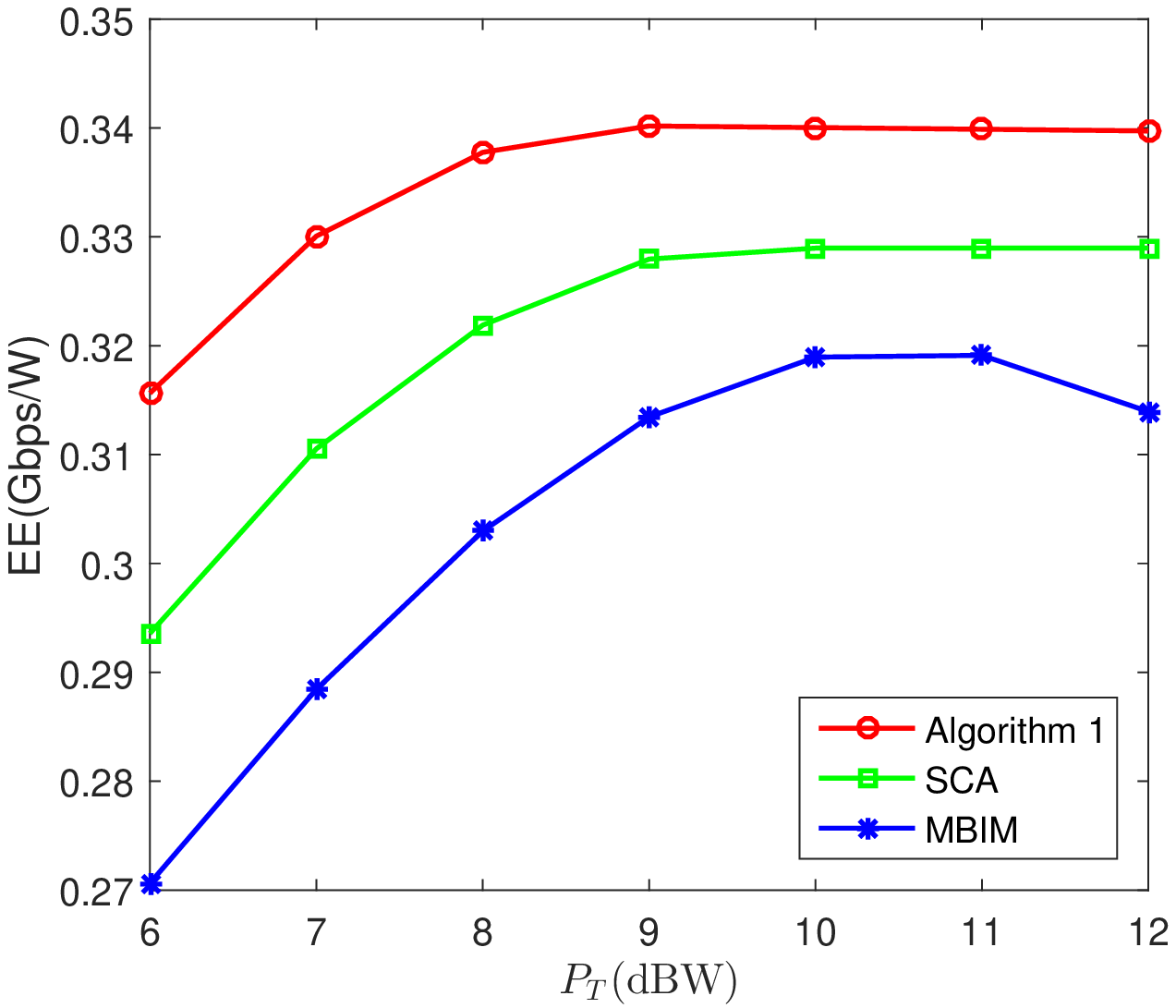}
\caption{EE comparisons for different $P_T$.}
\label{EE_PT}
\end{figure}

\begin{figure}[!t]
\centering
\includegraphics[width=90mm]{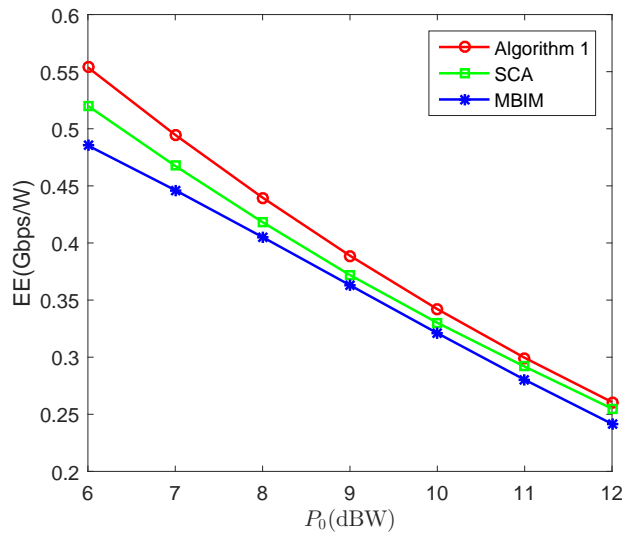}
\caption{EE comparisons for different $P_0$.}
\label{EE_P0}
\end{figure}

Fig. \ref{EE_PT} compares the EE of these three algorithms for different $P_T$, where we fix $P_0=10$dBW. When increasing $P_T$ from 6dBW to 10dBW, the EE of three algorithms all grows monotonically, implying that increasing the transmission power can effectively improve the EE performance. When further increasing $P_T$ from 10dBW to 12dBW, the EE of \textbf{Algorithm~1} and SCA keeps almost the same, indicating that merely increasing the transmission power cannot always improve the EE performance. To achieve the maximal EE, $P_T=10$dBW is enough for these two algorithms if there is no requirement of the achievable rate. Nevertheless, for the MBIM, further increasing $P_T$ from 10dBW to 12dBW results in the fall of the EE performance, which means the achievable rate grows more slowly than the transmission power and merely increasing the transmission power is sometimes nonbeneficial for the EE performance. We can observe that the maximal EE of 0.3392Gbps/W, 0.3327Gbps/W and 0.3241Gbps/W is achieved when $P_T=9$dBW, $10$dBW and $10$dBW for \textbf{Algorithm~1}, SCA and MBIM, respectively. Among the three algorithms, \textbf{Algorithm~1} is the best, owing to the fact that the two-step quadratic transformation for \textbf{Algorithm~1} can convert the original EEM problem of the MSS into an equivalent convex one while there is some performance loss during the approximated derivation for both SCA and MBIM.

Fig. \ref{EE_P0} compares the EE of these three algorithms for different $P_0$, where we fix $P_T=10$dBW. It can be seen that as $P_0$ increases, the EE performance of three algorithms all decreases monotonically, which implies that reducing the power consumption other than the transmission power of the satellite can effectively improve the EE.
%

\section{Conclusions}\label{sec.conclusion}
This letter have studied the EEM problem of the MSS under the constraints of total power and QoS. We have presented a two-step quadratic transformation method, where the original EEM problem can be converted to an equivalent convex one. Then an alternating optimization algorithm has been presented to solve it by iteratively performing two stages until a stop condition is satisfied. Simulation results have shown that it can fast converge and its performance is better than that of the SCA and MBIM. In the future, we will further study the efficient algorithms to improve the energy efficiency of satellite communications.

\bibliographystyle{IEEEtran}
\bibliography{IEEEabrv,IEEEexample}

\end{document}